\documentstyle[twoside,fleqn,espcrc2,epsfig]{article}
\def\lesssim{\mbox{ \raisebox{-1mm}{$\stackrel{<}{\sim}$} }}
\def\gtrsim{\mbox{ \raisebox{-1mm}{$\stackrel{>}{\sim}$} }}
\title{Axions from wall decay}
\author{S. Chang\address{
Department of Physics, University of Florida, Gainesville, FL 32611},
C. Hagmann\address{
Lawrence Livermore National Laboratory, Livermore, CA 94550}
and P. Sikivie$^{\small\rm a}$
}

\begin{document}
\begin{abstract}
We discuss the decay of axion walls bounded by strings 
and present numerical simulations of the decay process.  
In these simulations,
the decay happens immediately, in a time scale of order the
light travel time, and the average energy of the radiated axions
is $\langle \omega_a \rangle \simeq 7 m_a$ for $v_a/m_a\simeq 500$.  
$\langle \omega_a \rangle$ is found to increase approximately linearly
with $\ln(v_a/m_a)$. Extrapolation of this behaviour yields $\langle \omega_a
\rangle \simeq 60 m_a$ in axion models of interest.

\end{abstract}
\maketitle
\section{Introduction}
\label{sec:in}

The axion is the  quasi-Nambu-Goldstone boson associated with the spontaneous
breaking of the $U_{PQ}(1)$ symmetry which Peccei and Quinn postulated \cite{PQWW}.
Its zero temperature mass is given by:
\begin{equation}
m_a\simeq 6\cdot 10^{-6}\mbox{eV}\cdot N\cdot \left(\frac{10^{12}\mbox{GeV}}
{v_a}\right) \label{1.1}
\end{equation}
where $v_a$ is the magnitude of the vacuum expectation value that breaks
$U_{PQ}(1)$ and $N$ is a positive integer which describes the color anomaly of
$U_{PQ}(1)$.  
The axion owes its mass to non-perturbative QCD effects. At
temperatures high compared to QCD scale, 
these effects are suppressed and the
axion mass is negligible.
There is a critical time $t_1$, such that
$m_a(t_1) t_1 = 1$, when the axion mass effectively turns on.
The corresponding temperature $T_1 \simeq 1$ GeV.

In case that inflation occurs with reheat temperature higher than $v_a$,
axion strings form and radiate axions till $t_1$.   At
$t_1$ each string becomes the boundary of $N$ domain walls.  If $N=1$,
the network of walls bounded by strings is unstable 
and decays away. 
There are in this case three contributions to the axion cosmological
energy density:  
axions that were radiated by axion strings before $t_1$,
axions from vacuum realignment, and
axions that were produced in the decay of walls bounded by strings after
$t_1$. Here, we discuss the third of these contributions. 

That axions walls bounded by string decay predominantly into barely
relativistic axions was suggested in Ref.~\cite{Hagmann}.
D.~Lyth \cite{Lyth}
also discussed the wall decay contribution and emphasized the uncertainties
affecting it.  Nagasawa and Kasawaki \cite{Nagasawa} performed computer
simulations of the decay of walls bounded by string and obtained
$\langle \omega_a \rangle /m_a \simeq 3$ for the average energy of the radiated
axions.  Our simulations, presented in section 2, are similar to those of
Ref.~\cite{Nagasawa}
but done on larger lattices and for a
wider variety of initial conditions.  We obtain $\langle \omega_a \rangle / m_a
\simeq 7$ where $v_a/m_a\simeq 500$.  We attribute the difference between our result and that of
Ref.~\cite{Nagasawa} to the fact that we give angular momentum to the
collapsing walls.

The lattice sizes which are amenable to
present day computers are at any rate small compared to what one
would ideally wish.  
The lattice constant must be smaller than 
$\delta\equiv (\sqrt{\lambda} v_a)^{-1}$ to resolve the string
core.  On the other hand the lattice size must be larger than
$m_a^{-1}$ to contain at least one wall. Hence the lattice size in units
of the lattice constant must be of order
$\frac{10}{m_a\delta} \times \frac{10}{m_a \delta}$
or larger if the simulations are done in 2 dimensions.
Present day computers allow lattice sizes of order $4000\times 4000$, i.e.
$\frac{\sqrt{\lambda}v_a}{m_a}\sim 100$.  In axion models of interest
$\frac{\sqrt{\lambda}v_a}{m_a} \sim \frac{10^{12}{\rm GeV}}{10^{-5}{\rm eV}}
=10^{26}$.   
To remedy this shortcoming, we study the dependence of
$\langle \omega_a \rangle / m_a$ upon the $\frac{\sqrt{\lambda}v_a}{m_a}$ 
and find that it increases approximately
 as the logarithm of that quantity; see section 2.

The following toy model incorporates the qualitative features of interest: 
\begin{eqnarray}
{\cal L} &=&\frac{1}{2} \partial_{\mu} \phi^\dagger \partial^{\mu} \phi
-\frac{\lambda}{4}(\phi^\dagger\phi-v_a^2)^2 + \eta v_a\phi_1 \nonumber\\
&=&\frac{1}{2} \partial_{\mu} \phi_1 \partial^{\mu} \phi_1 +
\frac{1}{2}\partial_{\mu} \phi_2 \partial^{\mu} \phi_2 
\nonumber\\ &&-\frac{\lambda}{4}
(\phi_1^2+\phi_2^2 -v_a^2)^2 + \eta v_a\phi_1
\label{vi1}
\end{eqnarray}
where $\phi=\phi_1+ i\phi_2$ is a complex scalar field. When $\eta=0$,
this model has a $U(1)$ global symmetry under which
$\phi(x) \rightarrow e^{i\alpha} \phi(x)$. It is analogous to the $U_{PQ}(1)$
symmetry of Peccei and Quinn. It is spontaneously broken by the vacuum
expectation value $\langle \phi \rangle = v_a e^{i\alpha}$. The associated
Nambu-Goldstone boson is the axion. The last term in Eq.~(\ref{vi1})
represents the non-perturbative QCD effects that give the axion its mass
$m_a$. We have $m_a=\sqrt{\eta}$ for $v_a\gg m_a$.

When $\eta=0$, the model has global string solutions. A straight global
string along the $\hat{z}$-axis is the static configuration:
\begin{equation}
\phi(\vec{x})=v_a f(\rho) e^{i\theta}
\label{2.2}
\end{equation}
where $(z,\rho,\theta)$ are cylindrical coordinates and $f(\rho)$ is a
function which goes from zero at $\rho=0$ to one at $\rho=\infty$ over a
distance scale of order $\delta \equiv (\sqrt{\lambda} v_a)^{-1}$.
The energy per unit length of the global
string is
\begin{equation}
\mu \simeq  2\pi \int^L_\delta \rho d\rho \frac{1}{2} |\vec{\nabla}\phi|^2 =
\pi v_a^2 \ln (\sqrt{\lambda}v_aL)
\label{2.3}
\end{equation}
where $L$ is a long-distance cutoff. 
For a
network of strings with random directions, as would be present in the early
universe, $L$ is of order the distance between strings.

When $\eta\neq 0$, the model has domain walls. If $v_a\gg m_a$, 
the low energy effective Lagrangian is :
\begin{equation}
{\cal L}_a = \frac{1}{2}\partial_\mu a\partial^\mu a + m_a^2 v_a^2\left[\cos
\frac{a}{v_a}-1\right] .
\label{vi2}
\end{equation}
Its equation of motion has the well-known Sine-Gordon soliton solution :
\begin{equation}
\frac{a(y)}{v_a}= 2\pi n + 4 \tan^{-1}\exp(m_a y)
\label{2.5}
\end{equation}
where $y$ is the coordinate perpendicular to the wall and $n$ is any integer.
Eq.~(\ref{2.5}) describes a domain wall which interpolates between
neighboring vacua in the low energy effective theory (\ref{vi2}). 

The thickness of the wall is of order $m_a^{-1}$. The wall energy per unit
surface is $\sigma= 8 m_a v_a^2$ in the toy model.  In reality the structure
of axion domain walls is more complicated than in the toy model, mainly
because not only the axion field but also the neutral pion field is spatially
varying inside the wall \cite{Huang}. When this is taken into account, the
(zero temperature) wall energy/surface is found to be:
\begin{equation}
\sigma \simeq 4.2~f_\pi m_\pi f_a \simeq 9~m_a f_a^2\
\label{2.6}
\end{equation}
with $f_a \equiv v_a/N$.

\begin{figure}
\vspace{-2mm}
\centerline{
\includegraphics[width=7.5cm]{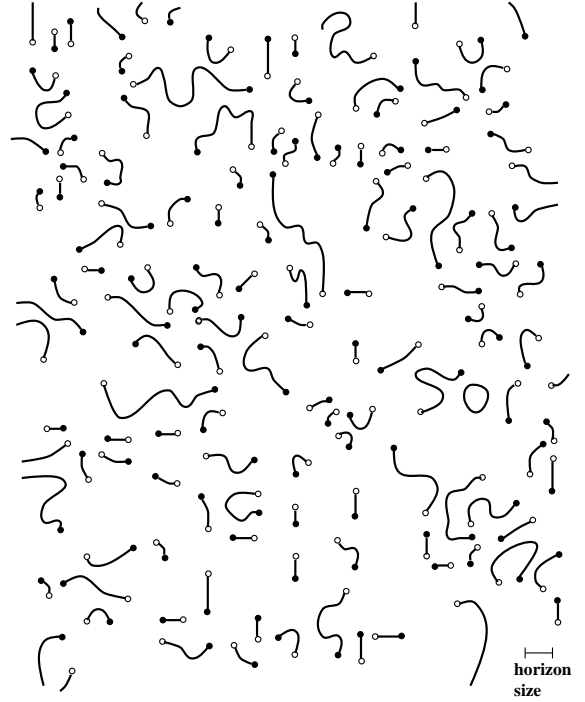}
}
\vspace{-9mm}
\caption{\small{A small cross-section of the universe near the QCD phase transition,
in $N=1$ axion models. Lines are domain walls, open (filled) circles are
upward (downward) going strings.}}
\vspace{-7mm}
\end{figure}

In the early universe, the strings are stuck in the plasma and are
stretched by the Hubble expansion.  However with time the plasma becomes
more dilute and below temperature \cite{Harari}
\begin{equation}
T_* \sim 2~10^7 \mbox{GeV} \left({v_a \over 10^{12} \mbox{GeV}}\right)^2
\label{2.7}
\end{equation}
the strings move freely and decay efficiently into axions. Because this decay
mechanism is very efficient, there is approximately one string per horizon
from temperature $T_*$ to temperature $T_1\simeq 1$ GeV when the axion
acquires mass. 

A cross-section of a finite but statistically
significant volume of the universe near time $t_1$ was simulated 
with the assumption that the axion field is randomly oriented from one horizon
to the next just before the axion mass turns on at time $t_1$ 
\cite{Paris,Chang}. Fig.~1 shows the result of this simulation.
The average density of walls at time $t_1$ predicted 
is approximately 2/3 per horizon volume at time $t_1$.

\section{Computer Simulations}
\label{sos}

\begin{figure}[b]
\vspace{-8mm}
\epsfxsize=80mm
\centerline{\epsfbox{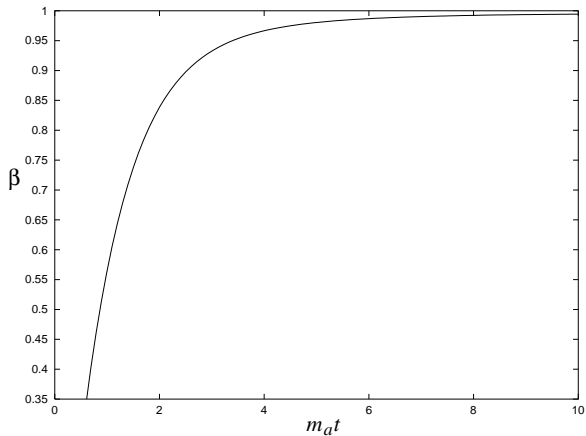}}
\vspace{-9mm}
\caption{\small{Speed of the string core as a function of time for the case
$1/m_a=400$, $\sqrt{\lambda}/m_a =10$, and $v =0$.}}
\label{lorentz}
\vspace{-3mm}
\end{figure}
We have carried out an extensive program of 2D numerical simulations of
domain walls bounded by strings. The Lagrangian (2) in finite difference
form is
\begin{eqnarray}
L&=&\sum_{\vec{n} }
\left\{ \frac{1}{2} \left[ \left(\dot{\phi}_1(\vec{n},t)\right)^2 +
\left(\dot{\phi}_2(\vec{n},t)\right)^2 \right]
\right.
\nonumber\\
&&- \sum^{2}_{j=1}\frac{1}{2}
\left[ \left(\phi_1(\vec{n}+\hat{j},t)
\left.\left.-\phi_1(\vec{n},t)\right)^2
\right.\right.\right.
\nonumber\\
&&\left.+ \left(\phi_2(\vec{n}+\hat{j},t)-\phi_2(\vec{n},t)\right)^2 \right]
\nonumber\\
&&
-\frac{1}{4} \lambda \left[\left(\phi_1(\vec{n},t)\right)^2
+\left(\phi_2(\vec{n},t)\right)^2 -1\right]^2 \nonumber\\ &&\left.
\hspace{-5mm}
\phantom{\frac{1}{2}}
+ \eta \left(\phi_1(\vec{n},t)-1\right)
\right\}
\end{eqnarray}
where $\vec{n}$ labels the sites In
these units $v_a=1$, the wall thickness is $1/m_a=1/\sqrt{\eta}$ and the core size
is $\delta=1/\sqrt{\lambda}$. The lattice constant is the unit of length.
In the continuum limit, the dynamics depends upon a single critical
parameter, $m_a \delta = m_a/\sqrt{\lambda}$.
Large two-dimensional grids $(\sim 4000\times 4000)$ were initialized
with a straight domain wall initially
at rest or with angular momentum. The initial domain wall 
was obtained by overrelaxation starting with the Sine-Gordon ansatz
$\phi_1+i\phi_2={\rm exp}(i\,{\rm tan^{-1}exp}(m_a y))$
inside a strip of length $D$
between the string and anti-string and the true vacuum $(\phi_1=1,\,\phi_2=0)$ outside.
The string and antistring
core  were 
approximated by $\phi_1+i\phi_2=-{\rm tanh}(.583\,r/\delta)\,
{\rm exp}(\mp i\theta )$ where $(r,\theta)$ are polar coordinates about
the core center,
and held fixed during relaxation. Stable domain walls were obtained for
$1/(\delta m_a )\gtrsim 3$.
A first-order in time and second order in space algorithm was used for
evolution with a
time step $dt=0.2$. The boundary conditions were periodic 
and the total energy was conserved to better than 1\%.
If the angular momentum was nonzero, then the time derivative $\dot{\phi}$
was obtained
by a finite difference over a small time step.

\begin{figure}
\epsfxsize=80mm
\centerline{\epsfbox{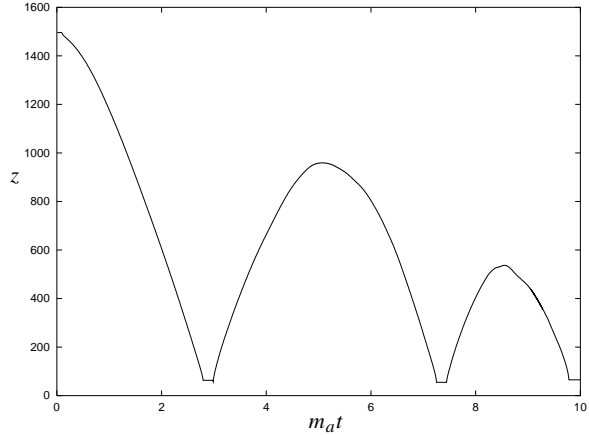}}
\vspace{-9mm}
\caption{\small{Position of the string or anti-string
core as a function of time for
$1/m_a=1000$, $\lambda = 0.0002, D=2896$, and $v=0$.  
The string and anti-string cores
have opposite $z$. They go through each other and oscillate with decreasing
amplitude.}}
\label{bounce}
\vspace{-7mm}
\end{figure}

The evolution of the domain wall was studied for various values
of $\sqrt{\lambda}/m_a$, the initial wall length $D$ and the initial
velocity $v$ of the strings in the direction transverse to the wall. 
Fig.~\ref{lorentz} shows the longitudinal velocity of the core as
a function of time for the case $m_a^{-1} = 400, \sqrt{\lambda}/m_a = 10$
and $v=0$.   An important feature is the Lorentz contraction of the core.
For reduced core sizes $\delta/\gamma_s \lesssim  5$,
where $\gamma_s$ is the  Lorentz factor associated with the speed of the
string core,
there is
'scraping' of the core on the lattice accompanied by emission of spurious
high frequency radiation. This artificial friction eventually balances the
wall tension and leads to a terminal velocity. In our simulations we always
ensured being in the continuum limit.

\begin{figure}
\epsfxsize=75mm
\centerline{\epsfbox{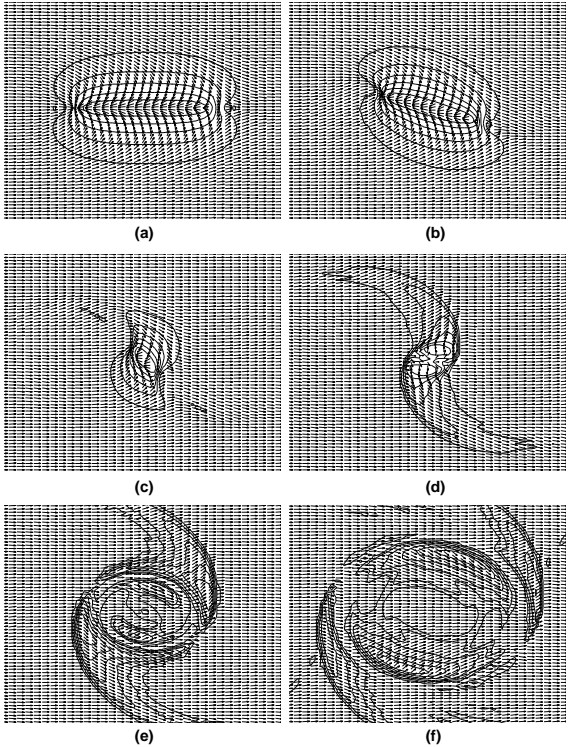}}
\vspace{-9mm}
\caption{\small{Decay of a wall at successive time intervals $\Delta t = 1.2/m_a$
for the case $m_a^{-1}=100$, $\lambda = 0.01$, $D=524$, and $v = 0.6$.}}
\vspace{-7mm}
\label{rot}
\end{figure}
For a domain wall without rotation ($v=0$), the string cores strike head-on
and go through each other.  Several oscillations of
decreasing magnitude generally occur before annihilation.  For
$\gamma_s \simeq 1$ the string and anti-string coalesce and annihilate one
another.  For $\gamma_s \gtrsim 2$, the strings go through each other and regenerate
another wall of reduced length.  The relative oscillation amplitude decreases
with decreasing collision velocity.  Fig.~\ref{bounce} shows the core position as
a function of time for the case
$m_a^{-1} = 1000, \sqrt{\lambda}/m_a = 14.3, D = 2896$.  

We also investigated the more generic case of a domain wall with angular
momentum.  The strings are similarly accelerated by the wall but string
and anti-string cores miss each other.  
The field is displayed in Fig.~\ref{rot} at six time steps for
the case $m_a^{-1} = 100, \lambda = 0.01, D = 524 $ and $v = 0.6$.  No
oscillation is observed.  All energy is converted into axion radiation during
a single collapse.  

\begin{figure}
\epsfxsize=77mm
\centerline{\epsfbox{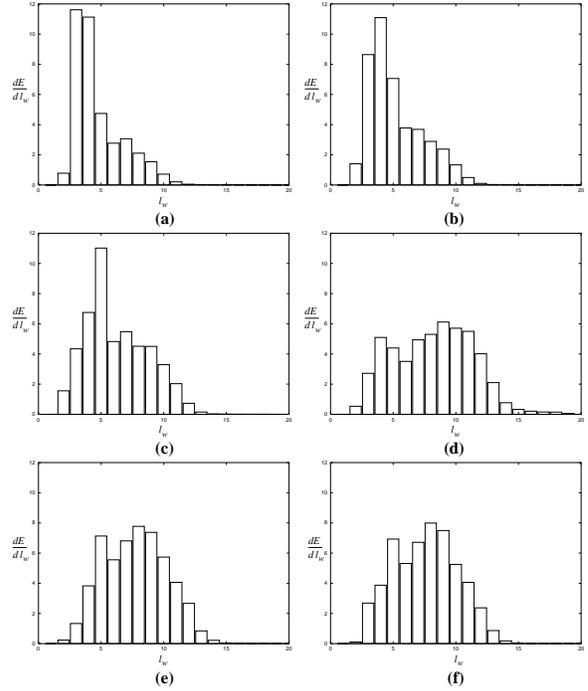}}
\vspace{-8mm}
\caption{\small{Energy spectrum at successive time intervals $\Delta t = 1/m_a$,
with  
$w_{\rm max}= \sqrt{8+m_a^2}$ and
$l_W=18 \ln(w/m_a)/\ln(w_{\rm max}/m_a)$ $+2$, 
for the case $m_a^{-1}=500$, $\lambda=0.0032$,
$D=2096$ and $v=0.25$.}}
\vspace{-7mm}
\label{spec}
\end{figure}
We performed spectrum analysis of the energy stored in the $\phi$ field
as a function of time using standard Fourier techniques.  The
two-dimensional Fourier transform is defined by
\begin{equation}
\tilde{f}(\vec{p})\!=\!\frac{1}{\sqrt{V}}\!\sum_{\vec{n}}\!
\exp\!\left[2i\pi\left(\frac{p_xn_x}{L_x}+\frac{p_yn_y}{L_y}\right)\right]\!
f(\vec{n})
\end{equation}
where $V\equiv L_x L_y$.  The dispersion law is:
\begin{equation}
\omega_p=\sqrt{2 \left(2- \cos\frac{2\pi p_x}{L_x}- \cos\frac{2\pi
p_y}{L_y}\right)+m_a^2}.
\end{equation}
Fig.~\ref{spec} shows the power spectrum $dE/d\ln \omega$ of the $\phi$ field
at various times during the decay of a rotating domain wall for the case
$m_a^{-1} = 500, \lambda = 0.0032, D = 2096$, and $v = 0.25$.  Initially,
the spectrum is dominated by small wavevectors, $k\sim m_a$.  Such a
spectrum is characteristic of the domain wall.  As the wall accelerates the
string, the spectrum hardens until it becomes roughly $1/k$ with a long
wavelength cutoff of order the wall thickness $1/m_a$ and a short wavelength
cutoff of order the reduced core size $\delta/\gamma_s$.  Such a spectrum is
characteristic of the moving string.  
\begin{figure}[h]
\vspace{-4mm}
\epsfxsize=75mm
\centerline{\epsfbox{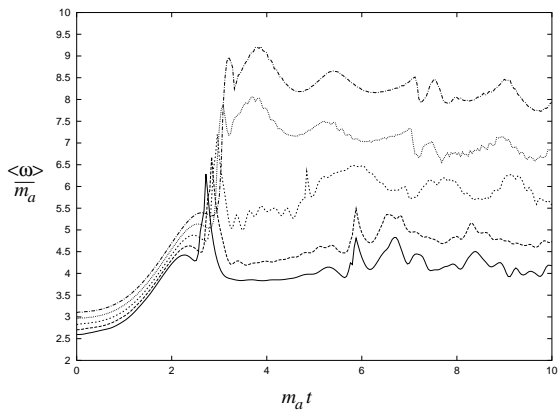}}
\vspace{-9mm}
\caption{\small{$\langle \omega \rangle$ as a function of time for $1/m_a=500$,
$D=2096$, $v=0.25$ and $\lambda=$ 0.0004 (solid), 0.0008 (long dash),
0.0016 (short dash), 0.0032 (dot) and 0.0064 (dot dash).  After the wall
has decayed, $\langle \omega \rangle$ is the average energy of radiated
axions.} }
\label{energy}
\vspace{-8mm}
\end{figure}

\begin{figure}[h]
\vspace{-1mm}
\epsfxsize=75mm
\centerline{\epsfbox{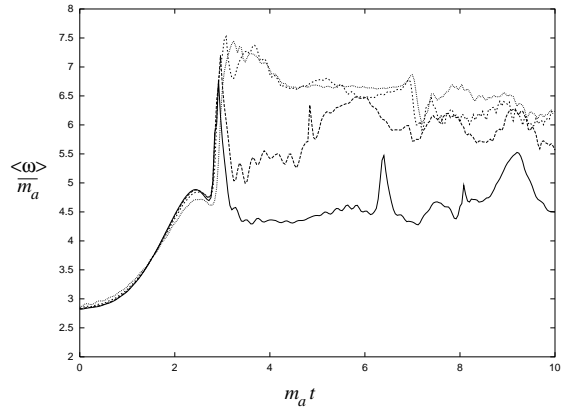}}
\vspace{-9mm}
\caption{\small{Same as in Fig.~\ref{energy} except $\lambda = 0.0016$, and $v=$ 0.15
(solid), 0.25 (long dash), 0.4 (short dash) and 0.6 (dot).}
}
\label{energy2}
\vspace{-7mm}
\end{figure}

Fig.~\ref{energy} shows the time evolution
of $\langle \omega \rangle/m_a$ for $m_a^{-1} = 500, v = 0.25, D = 2096$ and
various values of $\lambda$.  By
definition,
\begin{equation}
\langle \omega \rangle = \sum_{\vec{p}} E_{\vec{p}}/
\sum_{\vec{p}} \frac{E_{\vec{p}}}{\omega_{\vec{p}}}
\end{equation}
where $E_{\vec{p}}$ is the gradient and kinetic energy stored in mode
$\vec{p}$ of the field $\phi$.  After the domain wall has decayed into
axions, $\langle \omega \rangle = \langle \omega_a \rangle$.  Fig.~\ref{energy2}
shows the time evolution of $\langle \omega \rangle/m_a$ for
$m_a^{-1} = 500, \lambda = 0.0016, D = 2096$ and various values of $v$.

When the angular momentum is
low and the core size is big, the strings have one or more oscillations. 
In that case,
$\langle \omega_a \rangle/m_a \simeq 4$,  which is consistent with 
Ref.~\cite{Nagasawa}.  
We believe this
regime to be less relevant to wall decay in the early universe because it
seems unlikely that the angular momentum of a wall at the QCD epoch could
be small enough for the strings to oscillate.

In the more generic case when no oscillations occur,
$\langle \omega_a \rangle/m_a \simeq 7$.  Moreover, we find that
$\langle \omega_a \rangle/m_a$ depends upon the critical parameter
$\sqrt{\lambda}/m_a$, increasing approximately as the logarithm of that
quantity; see Fig.~6.  This is consistent with the time evolution of
the energy spectrum, described in Fig.~4.  For a domain wall,
$\langle \omega \rangle \sim m_a$ whereas for a moving string
$\langle \omega \rangle \sim m_a \ln (\sqrt{\lambda} v_a \gamma_s/m_a)$.
Since we find the decay to proceed in two steps: 1) the wall energy is
converted into string kinetic energy, and 2) the strings annihilate
without qualitative change in the spectrum, the average energy of
radiated axions is
$\langle \omega_a \rangle \sim m_a \ln (\sqrt{\lambda} v_a/m_a)$.
Assuming this is a correct description of the decay process for
$\sqrt{\lambda} v_a/m_a \sim 10^{26}$, then
$\langle \omega_a \rangle/m_a \sim 60$ in axion models of interest.

\section{Conclusions.}
\label{con}

We have presented results of our computer simulations of the motion and
decay of walls bounded by strings which appear
during the QCD phase transition in axion models with $N=1$ assuming the
axion field is not homogenized by inflation.
  In the simulations, the walls decay
immediately, i.e. in a time scale of order the light travel time.  The
simulations provide an estimate of the average energy of
the axions emitted in the decay of the walls: $\langle \omega_a \rangle \simeq
7 m_a$ when $\sqrt{\lambda}v_a/m_a \simeq 20$.

Because of restrictions on the available lattice sizes, the simulations are
for values of $v_a/m_a$ of order 100, whereas in axion models of interest
$v_a/m_a$ is of order $10^{26}$.  To address this problem, we have
investigated the dependence of $\langle \omega_a \rangle /m_a$ upon
$\sqrt{\lambda}v_a/m_a$ and found that it increases approximately as the
logarithm of this quantity.  This is because the decay process occurs in
two steps: 1) wall energy converts into moving string energy because the
wall accelerates the string, and the energy spectrum hardens accordingly;
2) the strings annihilate into axions without qualitative change in the
energy spectrum.  If this behaviour persists all the way to
$\sqrt{\lambda}v_a/m_a \sim 10^{26}$, then
$\langle \omega_a \rangle /m_a \sim 60$ for axion models of interest.

This has interesting consequences for the axion cosmology and the
cavity detector experiments of dark matter axion
as discussed in Ref.~\cite{Chang}.

\section*{Acknowledgments}
This research was supported in part by DOE grant DE-FG05-86ER40272 at the
University of Florida and by DOE grant W-7405-ENG-048 at Lawrence Livermore
National Laboratory.

\end{document}